\documentclass[%reprint,
superscriptaddress,
tightenlines,
nofootinbib,
twocolumn,
 amsmath,amssymb,
 aps,
]{revtex4-1}

\usepackage{amsmath}
\usepackage{amssymb}
\usepackage{bm}
\usepackage{color,graphicx}
\usepackage{colortbl}
\usepackage{slashed}
\usepackage{comment}
\usepackage{mathtools}
\usepackage{booktabs}
\usepackage{xcolor}
\newcounter{comment}

\begin{document}

\title{Benchmarks for a Global Extraction of Information from Deeply Virtual Exclusive Scattering  \begin{center} \vspace{0.5cm}
\includegraphics[scale=0.35]{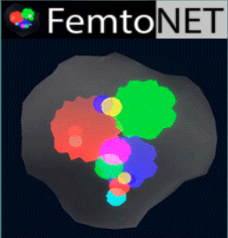}\end{center}}

\author{Manal Almaeen} 
\email{malma004@odu.edu}
\affiliation{Department of Computer Science, Old Dominion University, Norfolk, VA 23529, USA.}
\affiliation{Department of Computer Science, College of Computer and Information Sciences, Jouf University, Sakaka 72341, Aljouf, Saudi Arabia.}
\author{Jake Grigsby} 
\email{jcg6dn@virginia.edu}
\affiliation{Department of Computer Science, University of Virginia, Charlottesville, VA 22904, USA.}

\author{Joshua Hoskins} 
\email{jhoskins@jlab.org}
\affiliation{Department of Physics, University of Virginia, Charlottesville, VA 22904, USA.}

\author{Brandon Kriesten} 
\email{bkriesten@sura.org}
\affiliation{Center for Nuclear Femtography, Washington DC, 20005, USA.}

\author{Yaohang Li} 
\email{y7li@odu.edu}
\affiliation{Department of Computer Science, Old Dominion University, Norfolk, VA 23529, USA.}

\author{Huey-Wen Lin} 
\email{hwlin@pa.msu.edu}
\affiliation{Department of Physics, Michigan State University, East Lansing, MI 48824, USA.}
\affiliation{Department of Computational Mathematics,
  Science and Engineering, Michigan State University, East Lansing, MI 48824, USA}
  
\author{Simonetta Liuti} 
\email{sl4y@virginia.edu}
\affiliation{Department of Physics, University of Virginia, Charlottesville, VA 22904, USA.}

%\author{Sorawich Maichum} 
%\email{sm9cq@virginia.edu}
%\affiliation{Department of Physics, University of Virginia, Charlottesville, VA 22904, USA.}

%\affiliation{Laboratori Nazionali di Frascati, INFN, Frascati, Italy}

%\pacs{13.60.Hb, 13.40.Gp, 24.85.+p}

\begin{abstract}
We develop a framework to establish benchmarks 
%techniques 
for machine learning and deep neural networks analyses 
%to extract information from 
of exclusive scattering cross sections (FemtoNet). 
Within this framework we present  
%a Machine Learning based 
an extraction of Compton form factors  for deeply virtual Compton scattering from an unpolarized proton target.
%using a deep neural network . 
Critical to this effort is a study of the effects of physics constraints built into machine learning (ML) algorithms. We use the Bethe-Heitler process, which is the QED radiative background to  deeply virtual Compton scattering, to test our ML models and,  in particular, their ability to generalize information extracted from data. We then use these techniques on the full cross section and compare the results to analytic model calculations.
We propose a quantification technique, the random targets method, to begin understanding the separation of aleatoric and epistemic uncertainties as they are manifest in exclusive scattering analyses.
We propose a set of both physics driven and machine learning based benchmarks providing a stepping stone  towards applying explainable machine learning techniques with controllable uncertainties in a wide range of deeply virtual exclusive processes.
 \end{abstract}

\maketitle

\allowdisplaybreaks

%%%%%%%%%%%%%%%%%%%%%%
%%%  INTRODUCTION  %%%
%%%%%%%%%%%%%%%%%%%%%%
\section{Introduction}
\label{sec:intro}
\begin{figure}
    \centering
    \includegraphics[scale=0.75]{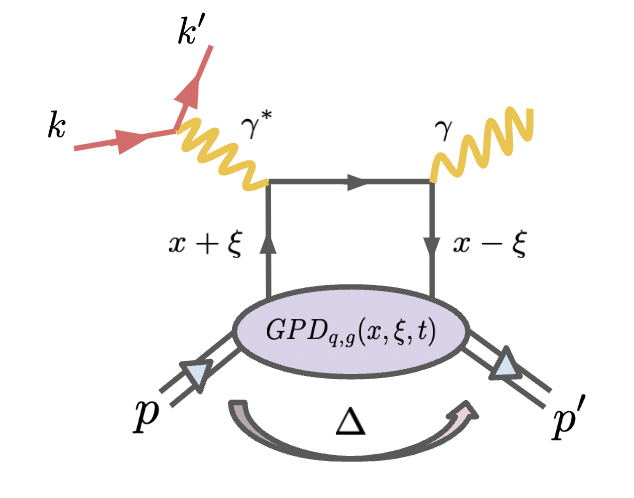}
    \caption{Depiction of Feynman diagram for the deeply virtual Compton scattering process parametrized through the GPDs.}
    \label{fig:feynman}
\end{figure}

Understanding the dynamical parton substructure of the nucleon is a fundamental goal of many nuclear physics experiments at both Jefferson Lab and at the upcoming electron ion collider (EIC) \cite{AbdulKhalek:2021gbh}. Deeply virtual exclusive scattering  (DVES) processes,  
\[ e p \rightarrow e' p' \gamma (M) \]
where either a high momentum photon ($\gamma$) or a meson ($M$), is detected along with the recoiling proton, can access the spatial distributions of quarks and gluons inside the proton through Fourier transformation of the process' matrix elements. 
Quantum chromodynamics (QCD) factorization theorems \cite{Ji:1997nk,Ji:1998xh} allow us to single out the correlation function for these processes, parametrized in terms of generalized parton distributions (GPDs). GPDs depend on the set of kinematic invariants $(Q^2, x_{Bj}, t, x)$, where $Q^2$, is the exchanged virtual photon four-momentum squared; $x_{Bj}$ is Bjorken $x$, proportional to the so-called skewness parameter $\xi$; the Mandelstam invariant $t$, gives the proton four-momentum transfer squared; $x$, the longitudinal momentum fraction carried by the struck parton, as shown in Figure \ref{fig:feynman} (we refer the reader to Refs.\cite{Diehl:2003ny,Belitsky:2005qn,Kumericki:2016ehc} for a review of the subject). 

Notwithstanding the validity of factorization theorems, an intrinsic difficulty affects the QCD analysis of DVES in that the cross section is parametrized in terms of form factors, known as Compton form factors (CFFs), which only depend on $(Q^2, x_{Bj}, t)$. 
%CFFs can, in principle be obtained directly from experiment.
%are measurable quantities
More precisely, GPDs -- the structure functions of the correlation function --  enter the CFFs -- the experimental observables -- only through convolutions over $x$, with Wilson coefficient functions determined in perturbative QCD (PQCD). The variable $x$, therefore, appears as an internal loop variable and is not directly observable. On the other side, PQCD evolution has been calculated at LO  in Refs.\cite{Ji:1996nm,Musatov:1999xp,Blumlein:1999sc,GolecBiernat:1998ja}, at %next-to-next-to-leading order (NNLO)
next-to-leading order (NLO) in \cite{Ji:1997nk,Ji:1998xh,Belitsky:1999hf,Freund:2001hd}, at NNLO in \cite{}, and it can be readily and easily implemented due to the collinearity of the partonic process. %These Compton form factors (CFFs) encode the information of hadron structure in complex experimental observables.
Among all DVES processes, deeply virtual Compton scattering (DVCS), where a photon is radiated from the initial proton, has been identified as the cleanest probe of GPDs. In this case one exploits the fact that the cross section for various polarization configurations contains an interference term with the QED Bethe-Heitler (BH) radiation process from the initial electron where 
%The interference term is directly proportional to linear combinations of 
the CFFs enter in linear combinations, making it easier to extract them from fits.  

Therefore, the goal of DVES data analyses is twofold: on one side CFFs need to be determined from the various cross sections and asymmetries. On the other, in a second step, GPDs need to be extracted from the CFF convolutions. 
An outstanding issue for these analyses is that for each electron-proton polarization configuration, {\it eight} CFFs enter simultaneously the cross section multiplied by multi-variable dependent kinematic coefficients. Owing to the challenges of this high-dimensional problem, results of the various analyses have been so far inconclusive.

The goal of our study is a precision extraction of CFFs, setting up the stage for an extraction of GPDs from the CFF convolutions. This task requires the implementation of new, advanced methods.
%, and the development of new specific approaches to the problem at hand. 
%Our advocacy for the use of machine learning (ML) techniques  is rooted  in this context.
Particularly, the ANN-based approaches developed so far need to be extended and develop into new specific machine learning (ML) techniques for the problem at hand.
Based in a statistical framework, ML applications in hadronic physics represent a new frontier for data analysis where many of the issues crucial for the extraction of CFFs and GPDs from data, including a systematic treatment of the different sources of error entering the analysis, can be properly addressed. 
Although ML algorithms have been used extensively to study high dimensional, ``raw", experimental data, the use in theory and phenomenology is still rather new \cite{Carleo:2019ptp}. 
Initial studies have been spearheaded by the NNPDF collaboration \cite{NNPDF:2021njg,NNPDF:2021uiq}, for the precision extraction of Parton Distribution Functions (PDFs) from a global analysis of high energy inclusive scattering data.
This is juxtaposed to parametric fitting procedures, including earlier ANN-based ones\cite{Ethier:2020way}, in which a functional form could bias the results when generalized to regions far outside of the data sets. 

%for future studies. Based in a statistical/mathematical framework, machine learning represents more than a ``black-box" technique for solving tricky problems. 
%
%An important distinction from other groups utilizing artificial intelligence techniques is that we are extracting CFFs directly from the data, not fitting parameters of a CFF parametrization to data. 
%
%This distinction allows us to directly propagate experimental error through our algorithms into extracted quantities and avoid bias from a parametric form. Moving beyond a supervised parameter fitting algorithm and utilizing the full power of ML to understand the physics in experimental observables presents an interesting question for future works.
%% 

The FemtoNet ML framework presented here utilizes physics-informed ML models with architectures that are specifically designed to automatically satisfy some of the physics constraints from the theory by limiting the predictions to only those allowed by the theoretical input, resulting in: {\it i)} less modeling error; {\it ii)} reduced demand on data points; {\it iii)} faster training;  {\it iv)} improved generalization. 
Within a collaborative effort involving physicists expert in both phenomenology and lattice QCD (LQCD), and ML specialists,
%and NN engineers, 
our goal has been  to develop and implement frontier techniques beyond standard supervised parameter fitting algorithms to exploit the full power of ML in understanding the physics information contained in experimental observables.
The more physics that is built into the architecture itself, the more power the model has in prediction and generalization.
Specifically, 
%to extract GPDs from experimental data, 
several theoretical constraints, reviewed in \cite{Kumericki:2016ehc}, are built in, including:
\begin{enumerate}
\item Cross section structure \cite{Kriesten:2019jep};
\item Lorentz invariance;
\footnote{Lorentz invariance results in the property of polynomiality of GPDs, namely the mathematically property of the Mellin moments of GPDs to be expressed in even powers of the skewness, $\xi$, multiplied by various $t$-dependent form factors} 
\item Positivity constraints \cite{Kumericki:2016ehc}; 
\item Forward kinematic limit,  defined by $\xi,t \rightarrow 0$, to PDFs, when applicable \cite{Kumericki:2016ehc}; 
\item $\Re e$-$\Im m$ connection of CFFs through dispersion relations \cite{Anikin:2007yh,Diehl:2007jb} with proper consideration of threshold effects \cite{Goldstein:2009ks};
\end{enumerate}
In order to apply some of these constraints, in particular, the property of Lorentz invariance/polynomiality, our analysis implements LQCD results on the GPD form factors \cite{Lin:2020rxa,Lin:2021brq}.

Most importantly, our {\it data-centric} approach affords us to accurately quantify 
%through rigorous study, 
the information that is contained in experiment through a more sophisticated definition of the uncertainty. 
A precise evaluation of the uncertainty in the extraction of CFFs from experimental data requires an almost exact overlap of the $(x_{Bj},t,Q^{2})$ kinematic coverage for all eight of the polarization observables entering the global analysis. Currently, this exact overlap is not found in experimental measurements. 
While a standard $\chi^{2}$  regression solves this issue by introducing a model dependent input, or functional forms to extract the CFFs, in the ML analysis, as described in what follows, we are able to 
%more sophisticated in that the 
separate the systematic error originating from the ML model itself 
%is distinct 
from the statistical error of the experimental data.
%This same problem exists for the extraction using ML models;
In our approach we have found that the final error on the predicted CFFs is almost exclusively from the ML model. We use {\it uncertainty quantification} techniques in order to quantify and control this systematic error. In particular, we adopted the uncertainty propagation technique described in \cite{gal2015dropout,gal2016dropout}, 
%in the ML literature, 
leaving a full exploration of uncertainty quantification methods to future work.
We remark that in the least squares fit, one does not define a model error, however, an imponderable systematic error is introduced, associated with the choice of the mathematical form of the parametrization. 
%This could work for PDFs where the functional forms are known to large extent, but it will not work for GPDs. 
%It will not work for precision studies where one is looking exactly for the exceptional behavior from a model norm. 

Several groups have already been proposing extractions of CFFs using various approaches differing both in their numerical and analytic components, {\it e.g.} using different formalism/approximations for the cross section, and/or different data sets and kinematic ranges  \cite{Cuic:2020iwt,moutarde2019unbiased}. It has, therefore, become mandatory to establish {\it benchmarks} to provide a solid common ground that will allow practitioners to conduct meaningful, quantitative comparisons, establishing whether results are compatible within error and identifying possible outliers.

In Section \ref{sec:benchmarks} we define a set of such benchmarks. As a step towards generalizing them for a comparison with other groups extractions, we subsequently conduct a test  using various choices of kinematics and other settings stemming from our analysis. The theoretical background for our analysis is presented in Section \ref{sec:background}, while the deep neural network for DVCS is described in Section \ref{sec:DNN_DVCS}. Our benchmark results are presented in Section \ref{sec:results}, and the error analysis in \ref{sec:error}. We finally provide an outlook and draw our conclusions in Section \ref{sec:outlooks}.

\section{Benchmarks for a machine learning based framework for Femtography}
\label{sec:benchmarks}

Critical to the analysis of deeply virtual exclusive reactions is a common set of benchmarking techniques for the extraction of information from cross sections and asymmetries. 
It is our goal to propose a set of standardized methods which will allow us to compare calculations of cross sections, 
%
%{\color{red} "It is our goal to propose a set standardized methods which..." Standards suggests there is, for instance, a tolerance outside of which comparisons are not valid. I think thi sis a better way to frame it.)}
set a precedent for the extraction of CFFs, and establish uncertainty quantification techniques for the propagation of statistical errors from the data. All of these benchmarks are necessary to conduct the following step of deconvoluting the GPDs from the extracted CFFs.

%We list below our standards for an analysis from which we can compare CFF extractions in the literature.  
We list below a standard set of metrics with which comparisons of CFFs in literature can be done.
%
 %{\color{red} The purposed metrics include:} 
 We propose a set of benchmarks for:  the physics description in which we compare the theoretical tensions of various cross section forms in the literature as well as features of the datasets used in the extraction;  the use of ML in the extraction of CFFs from cross sections and asymmetries including a set of standardized ML techniques;  the propagation of uncertainty from the error on the data to the error on the extracted CFF. 
In Section \ref{sec:outlooks} we give our perspective on new sources of information to be included in these benchmarks such as lattice QCD calculations, uncertainty quantification techniques from ML, and reinforcement learning methods in the extraction of GPDs.

\subsection{Physics Benchmarks}
We analyzed the various sources of theoretical uncertainty that are introduced in the physics description of DVES processes, and focused on DVCS off an unpolarized proton target. 
Differently from benchmarking techniques of PDFs that utilize a large set of processes to compare their extraction from experiment, for DVES there is not the same abundance of processes. We propose the use of the unpolarized cross section as a benchmark since it is the most well known, experimentally. Using the unpolarized cross section we can then compare CFF extractions at set kinematic points and compare to the data.

We singled out uncertainties due to the number and type of CFFs, and to the $Q^2$ dependence of the cross section. As we explain in detail in Sec.\ref{sec:background}, the cross section for unpolarized scattering has two components, for the unpolarized and polarized electron probe, respectively. Both components depends on all eight CFFs, multiplied by calculable coefficients. The analysis will therefore vary depending on the type and number of CFFs kept in the analysis. The $Q^2$ dependence is crucial since it originates from various sources: from kinematic terms, PQCD evolution, and higher twists.

%\textcolor{blue}{Compare cross sections with same formalism.}
A thorough understanding of the tensions between extracted quantities utilizing different data sets, at different kinematic points, generalized to regions outside of the data, and with different cross section formalism, will be necessary. In this section we will introduce some of the theoretical benchmark techniques that should be compared. 
However, we will utilize a single technique to show our results. 
A future publication with a more in depth study of the theory tensions and impacts on quantities such as the extracted error is warranted that will also address the question of what extent are differences in predictions due to different values used by each group.

%The aims were to establish the
%degree of compatibility and identify outliers amongst PDF sets, and to compare cross sections at the same αS values, thereby showing to what extent differences in predictions are
%due to the different αS values adopted by each group, rather than differences in the PDFs
%themselves. 

\subsection{Benchmarks for using ML}
Modern and deep learning techniques  have been applied to experimental nuclear physics  
%{\color{red}(Isn't the comparison between standard machine learning techniques (regression, KNN, random forests ...ect. and deep neural networks? I think "classical" ML models is a mischaracterization in terms of what I have heard used in the field. Suggestion: Modern ML and deep learning techniques ... or something like that)}.
Our recent study~\cite{Grigsby:2020auv} shows that deep ML models exhibit better flexibility in fitting the cross section curves compared to linear models, theory model, and Support Vector Regression (SVR). Standardized tools are available in the ML community to build deep neural network architectures. Evaluations of ML architectures and their associated hyperparameters, including the number of layers, the size of hidden nodes, activation functions, drop-out rates, loss functions, and gradient descent methods, are important to tune and optimize the ML models for global extraction of DVCS. In addition, data-centric analysis techniques, such feature selection and transformation, data augmentation, data synthetics, and data cleansing, can be applied to further enhance the training of the deep ML models. Moreover, well-known physics laws, such as cross section symmetries, energy and momentum conservation, and reduction of cross section through BH, can be incorporated into the deep learning neural network models, so called Physics-informed neural networks (PINNs), to tighten the problem space. Benchmarking the physics-informed strategies can provide useful guidelines to achieve more accurate fitting as well as better generalization of the extracted DVCS quantities. Furthermore, patterns reflecting the correlation between physics quantities are often captured in the latent space of the deep ML models when they are properly trained. Exploring and analyzing the learned patterns using data analysis techniques, such as dimension reduction and clustering, can further our understanding of the underlying physics and make the deep ML model explainable. 

% \begin{itemize}

%     \item{ML has been applied to experimental nuclear physics, we are taking this and applying it to nuclear theory, in particular exclusive phenomenology. Small datasets = novel ML techniques. Going beyond PDFs into GPDs (whole different beast)}
%     \item{Data centric analysis, not a fit of a parametric form. Extraction of numbers from data not fitting parameters to data.}
%     \item Machine learning is a statistical framework utilizing state of the art mathematical algorithms to solve complex problems.
%     \item Standardized techniques exist in the ML community to build and optimize neural network architectures. ML practitioners know these techniques and know how to utilize them to best solve individual problems. (eg. data augmentation, hyper-band algorithm)
%     \item A collaboration between both worlds will optimally benefit both the ML community and the physics community. Physics problems are extensive and difficult, making them interesting problems that require creative solutions. Physics informed neural networks (PINNs) to study trends of the data and discuss the errors of extracted quantities. (eq. cross section symmetries, reduction of cross section through BH)
%     \item We can address the dependence of the results on the neural network architecture through usage of various ML techniques. We showed in our last paper that DNNs are the best tool for analyzing DVCS data as compared to other computational techniques.
% \end{itemize}

\subsection{Benchmarks for Uncertainty Quantification}
A prerequisite for ML based analyses is that the underlying physics ground truth lies within the error bands of the predicted results or, in other words, that the underlying physics description is fully captured by the analysis. 
%{\color{red} I think know what this is trying to say but it needs to be rewritten in my opinion}. 
The estimated uncertainties also provide a way to evaluate the performance of various ML models.
There are four factors vital for the cause of uncertainty here: 
\begin{enumerate}
    \item The inherent statistical fluctuations in physics.
    \item The errors inherent to the measurement system.
    \item The errors in the ML models.
    \item The errors in the training procedure.
\end{enumerate}
%% THIS SHOULD BE EXPLAINED BEtter in a second round (SL)
%
% {\color{red}Aren't 3 \& 4 essentially the same thing. the model error is defined by the train procedure errors? Unless you mean the inherent limitation of the model given the parameter you have defined.}
Specifically, 1) and 2) are irreducible data-dependent uncertainties, which arise from the complexity of physics and experiments, while 3) and 4) are model-dependent uncertainty, which may be reduced by improving the ML models.

The predictive uncertainty of the deep ML models is of particular interest, which is the result of data- and model-dependent uncertainties propagated through the ML training process at different stages. Many uncertainty estimation methods for deep NNs can be applied here. For example, the Bayesian approximation method, such as Monte Carlo dropout~\cite{Gal2016} or Langevin dynamics~\cite{Welling2011} approximation, can be incorporated into the NN  architectures. Ensemble-based uncertainty quantification methods~\cite{Lakshminarayanan2017}, where an ensemble of deep ML models are independently trained, can be used to estimate the prediction confidence.

% {\color{blue} I feel like Subsections B \& C of this Section should be discussed after the model discussion. This is the theory benchmarking discussion after all. If nothing else it should be only very briefly discussed since your audience still doesn't have a good context of what was done yet.}
% We don't explore different uncertainty quantification techniques in this paper; we choose a specific method, dropout as a form of error estimation. Through a combination of dropout and data augmentation which allows us to estimate uncertainty on predicted quantities, a method we refer to as random targets.

% (rough outline/brain dump)
% The propagation of uncertainty in neural networks is notoriously difficult \cite{}. A neural network model can approximate any continuous function by mapping inputs to outputs through a series of layers; however, the way that the neural network does this is not straightforward. So unlike an analytic function where uncertainty propagation is straightforward, propagating uncertainty through a neural network and its non-linear transformations make this more difficult. 

%We utilize a method for uncertainty propagation through the application of a combination of dropout and data augmentation. This combo we use to propagate both the uncertainty from the data and the uncertainty inherent to the network.

%%%  BACKGROUND  %%%
\section{Background}
\label{sec:background}

%%%  DVCS THEORY  %%%
\subsection{DVCS Theory}
\label{sec:dvcs} 

The cross section for deeply virtual exclusive photoproduction, $ep \rightarrow e'  p' \gamma$, on a proton reads \cite{Diehl:2003ny,Belitsky:2005qn,Kumericki:2016ehc},
\begin{equation}
\label{eq:xs}
\frac{d^5\sigma}{d x_{Bj} d Q^2 d|t| d\varphi d \varphi_S } =
\Gamma \,  
\big|T\big|^2 \;
\end{equation}
where at leading order in $1/Q^{2}$, $T$ is given by the superposition of the amplitudes for DVCS and a competing background process Bethe-Heitler (BH),
\begin{equation}
|T|^2 = |T_{\rm BH} + T_{\rm DVCS}|^2
=|T_{\rm BH}|^2 + |T_{\rm DVCS}|^2 + \mathcal{I}\;
\label{eq:xsx}
\end{equation}
\begin{eqnarray}
\mathcal{I} & = & T_{BH}^{*} T_{DVCS}
+ T_{DVCS}^{*} T_{BH} 
\end{eqnarray}
where:
\begin{itemize}
\item{$Q^{2} = -(k-k')^{2}$ is the four momentum transfer squared between the initial ($e(k)$) and final ($e'(k')$) electrons}
\item{$x_{Bj}= Q^2/(2(pq))$ where $p$ and $q$ are the 4 vectors of the initial proton and virtual photon respectively.}
\item{$t = \Delta^{2} = (p'-p)^{2}$, the four momentum transfer squared between the initial and final state proton}
\item{$\varphi$ is the azimuthal angle between the lepton and hadron planes}
\item{$\varphi_S$, the azimuthal angle of the transverse proton spin vector, which we integrate over in our analysis since we are not considering scattering from a polarized target.}
\end{itemize}

Encoded in the amplitude, $T$, are the Compton form factors, which are convolutions of GPDs with perturbatively calculable Wilson coefficient functions. The CFF $\mathcal{F}$ where $F \in \{H,E\}$ is given as:

\begin{eqnarray}
\mathcal{F}^{q}(\xi,t) &=& \mathcal{C}(x,\xi;Q^{2}) \otimes F^{q}(x,\xi,t) \nonumber \\
&=&   e_{q}^{2}\int_{-1}^{+1}dx \Big[\frac{1}{\xi - x - i \epsilon} - \frac{1}{\xi + x - i\epsilon} \Big] F^{q}(x,\xi,t) \nonumber \\
\label{eq:cff}
\end{eqnarray}

\noindent and similarly for the polarized Compton form factors
\begin{eqnarray}
\widetilde{\mathcal{F}}^{q}(\xi,t) &=& \mathcal{C}(x,\xi;Q^{2}) \otimes \widetilde{F}^{q}(x,\xi,t) \nonumber \\
&=&   e_{q}^{2}\int_{-1}^{+1}dx \Big[\frac{1}{\xi - x - i 0} + \frac{1}{\xi + x - i0} \Big] \widetilde{F}^{q}(x,\xi,t) \nonumber \\
\label{eq:cff}
\end{eqnarray}

\noindent One can separate out the real and imaginary parts of the CFFs and sum over parton flavors for the total proton contribution that enters the cross section:

\begin{eqnarray}
\Re e \mathcal{F}(\xi,t) &=& e_{q}^{2}  \,P.V. \int_{-1}^{+1} dx \Big[\frac{1}{\xi - x} - \frac{1}{\xi + x } \Big] F^{q}(x,\xi,t) \nonumber \\ \\
\Im m \mathcal{F}(\xi,t) &=& \pi e_{q}^{2} \Big( F^{q}(\xi,\xi,t) - F^{q}(-\xi,\xi,t)  \Big)
\end{eqnarray}
where P.V. is the principal value integral. Similar expressions can be written for the polarized Compton form factors
\begin{eqnarray}
\Re e \widetilde{\mathcal{F}}(\xi,t) &=& e_{q}^{2}  \,P.V. \int_{-1}^{+1} dx \Big[\frac{1}{\xi - x} + \frac{1}{\xi + x } \Big] \widetilde{F}^{q}(x,\xi,t) \nonumber \\ \\
\Im m \widetilde{\mathcal{F}}(\xi,t) &=&  \pi e_{q}^{2} \Big( \widetilde{F}^{q}(\xi,\xi,t) + \widetilde{F}^{q}(-\xi,\xi,t)  \Big)
\end{eqnarray}
The $x$-dependence of the GPDs $H$ and $E$ directly describe physical properties of the nucleon such as partonic angular momentum in the proton through the angular momentum sum rule first written down by Ji \cite{Ji:1996ek}.
\begin{eqnarray}
J^{q/g} &=& \frac{1}{2}\int_{-1}^{+1}dx x(H^{q/g}(x,0,0) + E^{q/g}(x,0,0)) 
\end{eqnarray}
One can immediately see that the convolutions in the CFFs present a challenge for extracting GPDs from exclusive scattering experiments. The full $x$-dependence of the GPD is not directly accessible through any experimental observables; contrasted with inclusive scattering experiments from which the structure functions can be reconstructed through many measurements in values of $x_{Bj}$. The ultimate goal of nuclear femtography is to extract GPDs for nuclear imaging and understanding physical properties of the nucleon. We will break this down into a multi-step framework to go from cross section data to the physical properties encoded:
\begin{enumerate}
    \item One must be able to analyze trends in exclusive scattering data, understanding regions of high impact for experimental study.
    \item Extraction of Compton form factors from data with exploration of various error analysis techniques.
    \item Reconstruct the GPD using contributions from the lattice, and other theoretical constraints.
    \item Calculate physical quantities from the reconstructed GPDs necessary for femtography, including images.
\end{enumerate}

In this study, we will demonstrate the capabilities of ML algorithms in extracting Compton form factors from data including the propagation of experimental error on extracted quantities. We focus on unpolarized scattering data from Jefferson Lab 6 GeV and 12 GeV experiments. We will show the information that can be extracted from single polarization experimental observables. Our focus is the unpolarized scattering cross section.

The unpolarized scattering amplitude squared, $|T|^{2}$, can be separated into contributions from the Bethe-Heitler (BH) process, the pure DVCS process, and the interference between the two (INT). The total unpolarized cross section has all three contributions while the longitudinal beam and unpolarized target cross section (LU) only has a contribution from the interference term at leading twist. These various contributions can be written in terms of their CFF structure \cite{Kriesten:2019jep,Kriesten:2020apm,Kriesten:2020wcx},
\begin{widetext}
\begin{eqnarray}
|T_{UU}^{BH}|^{2} &=& \frac{\Gamma}{t}\Big[A_{UU}^{BH}\big(F_1^2 + \tau F_2^2 \big) + B_{UU}^{BH} \tau G_M^2(t) \Big] \\
|T_{UU}^{Int}|^{2} &=&   \frac{\Gamma}{Q^{2}t}\Big[A_{UU}^{\cal I}  \Re e \left(F_1 \mathcal{H} + \tau F_2  \mathcal{E} \right)   + B_{UU}^{\cal I}    G_M \Re e \left( \mathcal{H}+ \mathcal{E} \right)
 + C_{UU}^{\cal I}   
G_M \Re e\mathcal{ \widetilde{H}}  \Big] \\
|T_{LU}^{Int}|^{2} &=&   \frac{\Gamma}{Q^{2}t}\Big[A_{LU}^{\cal I}  \Im m \left(F_1 \mathcal{H} + \tau F_2  \mathcal{E} \right)   + B_{LU}^{\cal I}    G_M \Im m \left( \mathcal{H}+ \mathcal{E} \right)
 + C_{LU}^{\cal I}   
G_M \Im m\mathcal{ \widetilde{H}}  \Big] \\
|T_{UU}^{DVCS}|^{2} &=&  \frac{\Gamma}{Q^{2}}\frac{2}{1-\epsilon}\Big[ (1-\xi^2)\Big[  (\Re e {\cal H})^2 +  (\Im m {\cal H})^2 + (\Re e \widetilde{\cal H})^2 + (\Im m \widetilde{\cal H})^2 \Big] \nonumber \\
&+& \displaystyle\frac{t_o-t}{4M^2}  
\left[ (\Re e{\cal E})^2 + (\Im m{\cal E})^2  + 
 \, \xi^2 (\Re e\widetilde{\cal E})^2 +   \, \xi^2 (\Im m\widetilde{\cal E})^2 \right] \nonumber \\
&-&  2\xi^{2} \,\left( \Re e {\cal H} \, \Re e {\cal E } + \Im m {\cal H} \Im m{\cal E } +  \Re e \widetilde{\cal H} \,  \Re e\widetilde{\cal E } + \Im m \widetilde{\cal H} \Im m \widetilde{\cal E }   \right) \Big] 
\end{eqnarray}
\end{widetext}
Notice that the dimensions of the cross sections are all $(1/$GeV$^{2})$ where powers of GeV$^{2}$ can be hidden in the coefficients $A, B, C$, which are functions of $(Q^2, x_{Bj}, t)$, and of the initial electron energy, $k_o=E_e$ and $\varphi$. Detailed calculations of these cross sections including the calculable coefficients and a description of their dimensions can be found in \cite{Kriesten:2019jep,Kriesten:2020wcx}. It should be noted that other calculations of the cross section calculations exist in the literature (BKM '01 \cite{Belitsky:2001ns}, BKM '10 \cite{Belitsky:2010jw}, Braun et. al. \cite{Braun:2012hq}, Guo et al \cite{Guo:2021gru}) the differences between the formalism is discussed in \cite{Kriesten:2020wcx}.

In our analysis we will consider available fixed target data for an unpolarized target described by the set of observables: 
\[E_e, Q^2, x_{Bj}, t, \varphi.\] 
%
%\noindent  where $E_e$ is the energy of the initial electron. 
\noindent The unpolarized proton data represent a substantial percentage of the total DVCS dataset in terms of number of data points. It is important for this analysis to point out that the cross section has few points in the kinematic variables with most of the points in the azimuthal angle $\varphi$. This presents an interesting challenge for ML algorithms for generalization. We find that the asymmetry in experimental cross section for $\phi > \pi$ versus $\phi < \pi$, due to experimental uncertainty, presents a challenge for ML algorithms to learn the inherent symmetry property. It is through the application of symmetry constraints in the prediction of the cross section that allows the ML algorithm to predict more accurately and generalize outside of the data. We emphasize this concept in our analysis; however it should be noted that this point was first discussed in \cite{Grigsby:2020auv}.

%%%  NN DESCRIPTION  %%%
\subsection{Deep Neural Networks}
% Deep neural networks (DNNs) are statistical tools used to represent complex patterns and functional forms of data through layers of simple computational elements called neurons. 

% % In this paper, we continue our analysis of supervised DNN methods to extract CFFs from the data. First, we include some background on DNNs and Deep Learning.

Deep Neural Networks (DNNs) are a machine learning tool for function approximation that contain layers of non-linear transformations mapping input data to predictions. This paper will focus on multilayer perceptrons (MLPs), which are a standard approach for vector input data (e.g., arrays of kinematic data). MLPs pass their input data through a sequence of affine linear transformations - also known as ``fully connected" (FC) layers - with non-linear \textit{activation functions} applied element-wise between layers. The \textit{parameters} of the $j$th layer with input size $j_{\text{in}}$ and output size $j_{\textit{out}}$ are the matrix $W_j \in \mathbb{R}^{j_{\text{out}} \times j_{\text{in}}}$ and bias vector $b_j \in \mathbb{R}^{j_{\text{out}}}$. Layer $j$ projects its input $x_{j} \in \mathbb{R}^{j_{\text{in}}}$ to $x_{j+1} \in \mathbb{R}^{j_{\text{out}}}$ with a linear transformation followed by a non-linear activation function $z_j$:
\begin{eqnarray}
x_{j+1} = z_j(W_{j}x_{j} + b_j)
\end{eqnarray}
A common choice of activation function is the rectified linear unit (ReLU)~\cite{relu} $z_j(x) := \text{max}(x, 0)$ and its variants such as Leaky RELU~\cite{Maas2013RectifierNI} because all of their advantages like solving the vanishing Gradient  problem. Each layer can be thought of as a network where the coefficients of $W_j$ represent the strength of the connections between between $j_{\text{in}}$ input neurons and $j_{\text{out}}$ output neurons, and the bias and nonlinearity represent a minimum threshold for neuron ``activation." The full MLP, $f$, can be written as a composition of $k$ layers:

% DNNs are multilayered perceptrons constructed of a series of layers in which each layer is composed of a sequence of nodes. These nodes encode the previous layer's output followed by a nonlinear activation function which allows for more robust learning. One can write this sequence of layers as a functional composition

\begin{eqnarray}
f_{\theta}(x) &=& z_k(W_{k}(z_{k-1}(W_{k-1}(\dots (z_0(W_{0}x+b_{0})\dots\nonumber \\
&&+b_{k-1}))+b_{k}) 
\end{eqnarray}
\noindent where $\theta$ denotes the set of all parameters $\{(W_j, b_j) | j \leq k\}$ that make up the DNN. Ideally, deep networks let each layer learn a slightly improved representation of the input data, until accurate predictions can be made by a standard linear model in the final layer $k$. The choice of activation functions $z$, the number of layers $k$, and the size of each $W$ and $b$ parameter creates a network ``architecture." 

DNNs are trained to minimize a \textit{loss function}, $\mathcal{L}$, that represents the accuracy of their predictions. A common loss function for regression problems is  mean squared error (MSE). Given a dataset of $N$ pairs of inputs and desired outputs $\{x^{(i)}, y^{(i)}\}_{i=0}^{N}$, we compute the loss function on a batch of $B$ samples using our current parameters $\theta_t$. We can then differentiate the loss function with respect to $\theta_t$ and use the gradient vector to shift our parameters in a direction that has lower prediction error: 
\begin{eqnarray}
\label{sgd}
    \mathcal{L}_{\theta_t} &=& \frac{1}{2}(y - f_{\theta_t}(x))^2 \\
    \theta_{t+1} &=& \theta_t - \alpha \nabla_{\theta_t}\frac{1}{B}\sum_{i=0}^{B}\mathcal{L}_{\theta_t}(x^{(i)}, y^{(i)}) 
\end{eqnarray}
\noindent where $\alpha$ is a small scalar known as the \textit{learning rate}. In practice, it is common to treat $\alpha$ as a vector of size $|\theta_t|$ and adjust each element with various heuristics across multiple updates \cite{kingma2014adam}.

% \noindent where we have used the convention of $k$ the number of layers in the neural network architecture and the parameters $\theta$ include the nonlinear activation function $z_i$, the weights of the network $W_i$, and the biases $b_i$ at the $i$th layer. The weights and biases form the parameter space of the neural network. These parameters as well as the total size of the network layers (indicated by the number of nodes in each layer) make up the model's architecture. These choices can be optimized for peak performance of the network given some metric of error on the predictions. 

DNNs with thousands or millions of parameters can risk over-optimizing on (or ``memorizing") training inputs at the expense of their ability to generalize to unseen data. This problem is known as overfitting, and can be addressed by \textit{regularizing} the network during training. We choose a regularization technique called dropout \cite{srivastava2014dropout}, where elements of a layer's weight matrix $W$ are zeroed with probability $p$ by a binary mask that is randomly generated before each training step. Dropout prevents a network from becoming overly reliant on the connections between specific neurons. Dropout also allows us to estimate the systematic error of the neural network \cite{gal2015dropout}, where the distribution of a number of predictions with different dropout masks can approximate variational inference. Another approach reduces the network's ability to overfit to the training data by constraining the parameters to be close to the origin; this can be done by adding the L2 norm $||\theta_t||_2$ to the loss function (Eq. \eqref{sgd}) with coefficient $\lambda$.

%  \textcolor{orange}{where the
% dropout layers are formulated as Bernoulli distributed random
% variables and the neural network training with dropout layers can be approximated as performing variational inference.} As the neurons are systematically passed through the binary dropout mask, the output changes during each calculation of the predictions. As the number of predictions grow, we can treat the output of the network as a statistical ensemble and thus take the mean and the variance of the output set. This is what we treat as our predictive error.
%\JGNOTE{I think everything after this point isn't background and probably belongs in a dedicated method section.}

\section{Physics Constrained Deep Neural Networks for DVCS Cross Sections}
\label{sec:DNN_DVCS}
The use of DNNs has solved many complex problems in nuclear physics \cite{YasirSurvey2021,Almaeen2021}. %\JGNOTE{if this isn't a double-blind review we might as well be more clear that we are continuing our prior work}
It was shown in \cite{Grigsby:2020auv} that DNNs are an effective approach to DVCS data analysis that outperforms simpler machine learning baselines. Typically, training DNNs models require a large size of data that might not always be available or possible to obtain in physics applications. As an alternative, DNNs models can be trained using additional knowledge gained by enforcing some physics laws for better accuracy and generalization \cite{physics_informed}. Also, purely data-driven models may perfectly fit observations, but predictions might be inconsistent, as a result of extrapolation or observational biases that may lead to poor generalization 
performance. Thus, there is an imperative need for combining fundamental physical rules and domain knowledge by “teaching” machine learning  models physical laws, which can, in turn, provide “informative priors”. A recent example that follows this new learning technique is the family of “physics-informed neural networks” (PINNs) introduced in~\cite{raissi2019physics}. The major motivation for implementing these methods is that such prior knowledge or constraints can reduce the parameter space of the neural network to physics feasible space and yield more interpretable machine learning methods that are robust in the presence of imperfect data  such as missing values or outliers, and can provide  precise and physically consistent predictions, even for generalization tasks.

For deep learning models of exclusive experiments we introduce the following physics knowledge and constraints:
\begin{itemize}
    \item Error bars
    \item Parity Invariance 
    \end{itemize}
In a future development we plan to include a whole new set of constraints including from LQCD. 

\subsubsection{Model Architecture}

\begin{figure}[h]
\includegraphics[height=3.5cm, width=\linewidth]{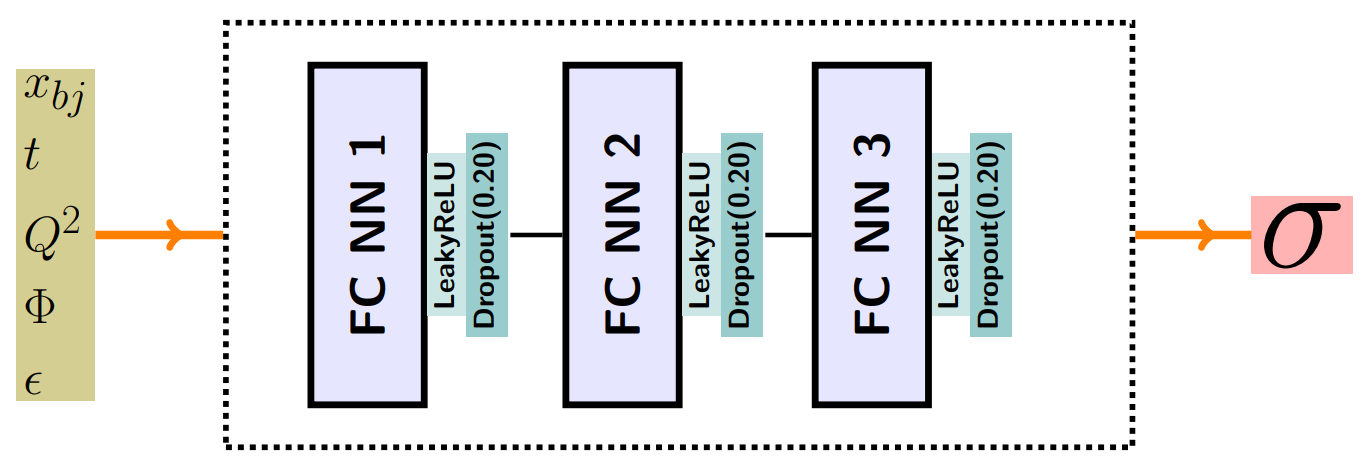}
\caption{Neural network architecture that takes in input variables $x_{Bj}, t, Q^{2}, \phi, \epsilon$ and outputs the cross section. Each fully connected layer in this network architecture is followed by a nonlinear activation function (Leaky RELU) and a dropout mask with $p = 0.2$. }
\label{fig:Arch}
\end{figure}

Often when constructing a machine learning model, there will be presented several design choices to define the architecture as we do not know in advance what the optimal architecture should be for a given model. Basically, training machine learning models involves two types of parameters. First, trainable parameters that are learned by the algorithm through the learning process such as the weights and biases of a neural network. Second, the hyperparameters that the learning algorithm will utilize to learn the optimal parameters that accurately map the input features to the targets. The hyperparameters such as the number of layers, number of neurons, and the learning rate,  need to be manually tuned before launching the training process. The hyperparameters tuning is an essential process of searching for an optimal set of hyperparameters that can leverage the highest performance of our model. Finding the optimal hyperparameters manually is a challenging task as there can be numerous hyperparameters. Therefore, we adopt Keras tuner \cite{omalley2019kerastuner}, an open-source library to conveniently define our architecture and perform hyperparameter tuning. The Keras tuner workflow as follows,  first, a tuner is specified to determine the hyperparameter combinations that should be tested. Second, the library search algorithm such as Bayesian optimization executes the iteration loop to evaluate a particular number of hyperparameter combinations. Third, evaluation is performed by calculating the loss of the trained model on a held-out validation set.  Finally, the optimal hyperparameter combination in terms of validation loss is retrieved to be tested on a held-out test set. The optimal hyperparameter is utilized as the final hyperparameter for our model architecture illustrated in Figure~\ref{fig:Arch}. Our optimum network architecture composed of an input layer whose nodes denote the array of the kinematic [$x_{bj}$, $t$, $Q^2$, $E_b$, $\Phi$], followed by three fully-connected layers with 1024 neurons activated by a Leaky ReLU function.The output layer represents our target cross section. The network is regularized by an L2-norm penalty and a dropout rate of 0.2 to prevent overfitting.
Based on the hyperparameter search algorithm, a shallow and wider network can have better performance than a deeper network with less number of neurons.

 Our model is optimized using a joint of three loss functions: The first and basic loss function is the mean squared error (MSE) to minimize the error between the predicted and observed cross sections. The second loss is to ensure that our predicted cross sections are symmetric. Lastly, a loss function to ensure the start and the end of the predicted data on a range of $\phi$ are matching. 

%%%%%
% \begin{figure}[h]
% \includegraphics[height=4.2cm]{figures/tuner_2.pdf}
% \caption{Hyperparameter tuning process}
% \label{fig:tuner}
% \end{figure}

% \begin{figure}
%     \centering
%     \includegraphics[height=4.2cm, width=\linewidth]{figures/tuner_2.pdf}
%     \caption{Hyperparameter tuning process.}
%     \label{fig:tuner}
% \end{figure}

%%%%

\subsubsection{Parity Invariance}

 The symmetry constraints on the cross section can be applied because there is no \textit{``phase changing"} term that can enter the unpolarized cross section through physics constraints of parity. Any parity violating terms related to beyond the standard model physics we treat as part of the extracted error as compared to the size of the experimental statistical error. Studies for beyond the standard model physics that can be extracted from DVCS data we leave to a future study. We then deem the asymmetry in the $\phi$ dependence in the unpolarized cross section as due to the systematic uncertainties in the experimental data taking. With many points in $\phi$ the data is more sensitive to these type of systematic errors. We also note that taking many data points in $\phi$ is based on a harmonics analysis for extracting CFFs which we are not utilizing for our extraction method. It has been demonstrated in \cite{Kriesten:2020apm} that there are alternative approaches for extracting quantities related to angular momentum beyond the harmonics description.

To incorporate angular symmetries into the machine learning model, we use additional loss function terms
$$ \|f(x_{Bj}, t, Q^{2}, \phi, \epsilon) - f(x_{Bj}, t, Q^{2}, -\phi, \epsilon)\|$$
and
$$ \|f(x_{Bj}, t, Q^{2}, \phi, \epsilon) - f(x_{Bj}, t, Q^{2}, \pi-\phi, \epsilon)\|,$$
measuring the unpolarized angular symmetry and polarized angular symmetry, respectively.

The incorporation of the additional loss function term is an example of the physics constraints that we propose to build into the architecture of the neural networks. It has been shown that such neural networks~\cite{Karniadakis2021-pa, RAISSI2019686} can improve the performance and generalizability of neural network predictions. Further physics inputs such as input from lattice QCD calculations, can additionally be built into neural network architectures at different steps of the analysis. 

\subsubsection{Data Augmentation}

Deep learning networks have made extraordinary progress in many scientific applications. However, these networks perform more effectively as there is a large amount of data available to train on. For example, text-based models have performed significantly better because of the release of a trillion-word corpus by Google \cite{35179}.
The reason is that training on large datasets can avoid overfitting where a network learns a function with very high variance such as to perfectly fit the training data but do not generalize well on the unseen data. There are several proposed methods to reduce the problem of  overfitting such as dropout \cite{JMLR:v15:srivastava14a}, batch normalization \cite{pmlr-v37-ioffe15}, transfer learning \cite{Weiss2016ASO,Shao2015TransferLF} and pretraining \cite{JMLR:v11:erhan10a}. In addition to these methods is the data augmentation that approaches overfitting from the root of the problem, the training dataset. This is performed under the assumption that additional data and information can be obtained from the original dataset through augmentations. Data Augmentation involves a set of techniques \cite{Shorten2019ASO} that increase the quality and size of training datasets so that more powerful deep learning models can be built.

\begin{figure}[h]
\includegraphics[width=\linewidth]{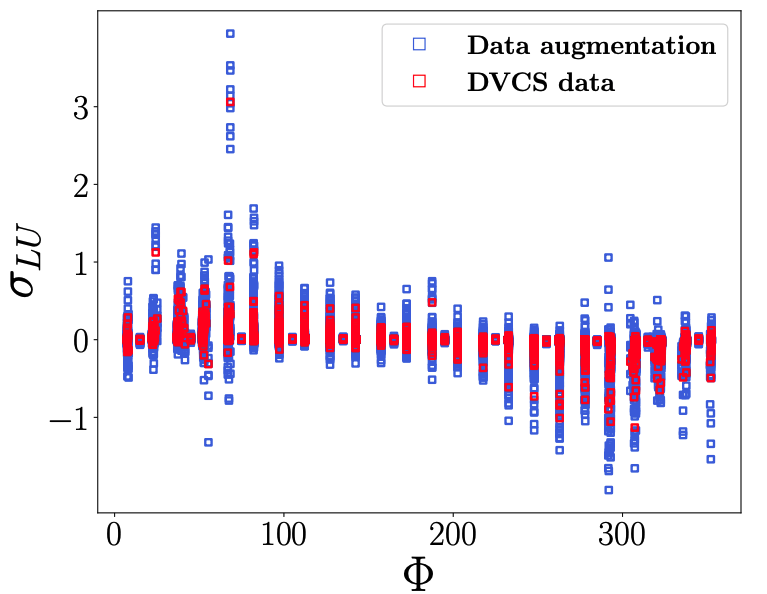}
\includegraphics[width=\linewidth]{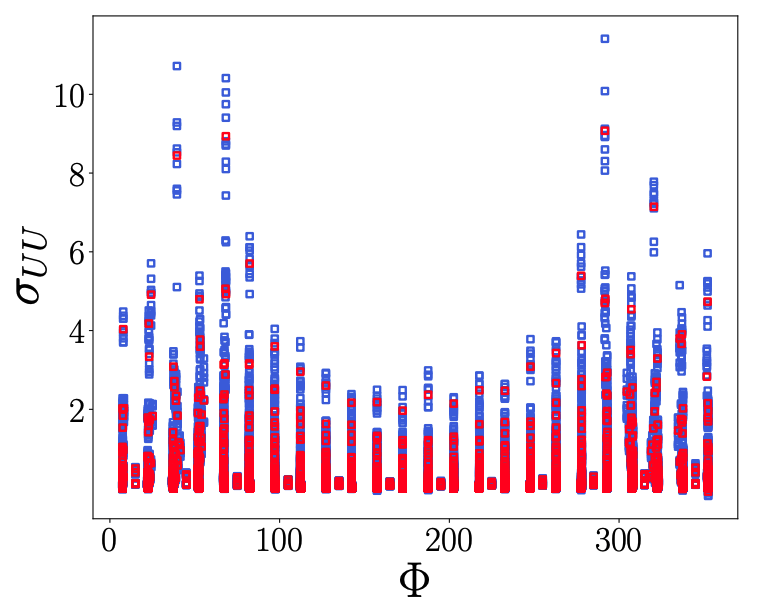}
\caption{Data augmentation on unpolarized (top) and polarized DVCS data(bottom).}
\label{fig:data_aug}
\end{figure}

We use data augmentation to increase the size of our DVCS data by adding slightly modified copies of our existing data. Since we have only 3,862 unpolarized and 3,884 polarized data points, data augmentation seems to be essential to apply in order to improve the model accuracy. Each DVCS data point includes a mean and a standard error indicating its uncertainty. The standard errors varies across data points. We increase the size of the data 10 times for each the unpolarized and polarized data by sampling the cross section using the statistical error $\Delta\sigma$. Data augmentation is applied on the unpolarized and polarized data independently as shown in Figure~\ref{fig:data_aug}. There are two main advantages of data argumentation. Firstly, more data samples become available for training. Secondly and more importantly, the trained neural network is informed with the uncertainty of the training samples, where more model flexibility is allowed on the sample points with larger standard errors and vice versa. As a result, the neural network trained with augmented samples yield better accuracy and generalization.

Including the information contained in the experimental standard errors into the architecture of the network through data augmentation. We are treating the experimental errors as additional sources of physics information, errors are not a nuisance but rather a source of information. This is another example of building physics into the architecture of the network and utilizes standard neural network practices.
%  \begin{figure}[http]
% \includegraphics[width=\linewidth]{figures/data_aug.png}
% \caption{Data augmentation on unpolarized (left) and polarized DVCS data(right).}
% \label{fig:data_aug}
% \end{figure}

%%%  RESULTS  %%%
\section{Results}
\label{sec:results}

%%%  CROSS SECTION GENERALIZATION  %%%
\subsection{Cross Section Generalization}
\begin{figure}
\includegraphics[width=\linewidth]{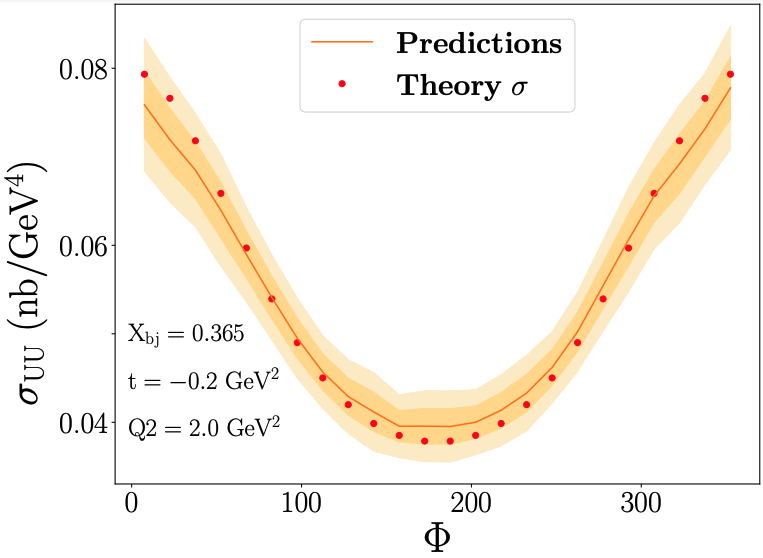}
\includegraphics[width=\linewidth]{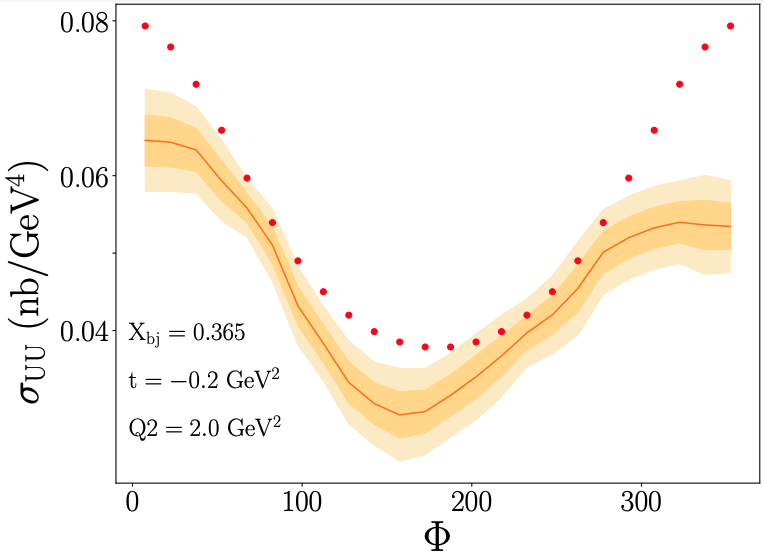}
\caption{DVCS extrapolation on kinematics outside of the range covered in experiment at the kinematic point $x_{Bj} = 0.365$, $t = -0.2$ GeV$^{2}$, $Q^{2} = 2$ GeV$^{2}$, and $E_{b} = 5.75$ GeV. (\textit{Top})ML Model with Angle Symmetric Constraints  (\textit{Bottom}) ML Model without Angle Symmetric Constraints.}
\label{fig:DVCS}
\end{figure}

In this section we show the results of our study of incorporating physics into the architecture of a deep learning model trained to predict DVCS cross sections and how this improves both performance and generalization capabilities. We test our model on the exactly calculated Bethe-Heitler process in order to calibrate our analysis. Our results on the predicted cross section in an extrapolated region of kinematics outside of the range covered in experiments are shown in Figure \ref{fig:DVCS}.

%\begin{figure}[h]
%     \begin{subfigure}[b]{0.49\textwidth}
%        \includegraphics[width=0.8\textwidth]{figures/constraint.png}
%         \caption{ML Model with Angle Symmetric Constraints}
%     \end{subfigure}
%     \hfill
%     \begin{subfigure}[b]{0.49\textwidth}
%        \includegraphics[width=0.8\textwidth]{figures/noconstraint.png}
%         \caption{ML Model without Angle Symmetric Constraints}
%     \end{subfigure}
%     \caption{DVCS extrapolation on kinematics outside of the range covered in experiment.}
%    \label{fig:DVCS}
%\end{figure}

A single cross section data point represents a point in a multi-dimensional kinematic phase space. In the context of local extractions, this point contains all of the information one can extract. However, it is necessary for a global analysis to study trends in this higher-dimensional space. In this sense, a group of $N$ data points represents more than $N$ disjoint pieces of information but rather a region of trends and patterns that can inform predictions in kinematic regions outside of the data. This includes correlations between data points that are not taken into account/cannot be taken into account in point-by-point extractions.

Neural networks learn complex representations of data while extracting trends that can be generalized to regions beyond the data. Generalization represents a unique opportunity to squeeze as much information from DVCS experiments as possible while quantifying learned uncertainties. These generalization capabilities will allow us to inform experiments of worthwhile, high impact measurements in the extraction of Compton form factors.

\begin{figure}
\includegraphics[width=0.4\textwidth]{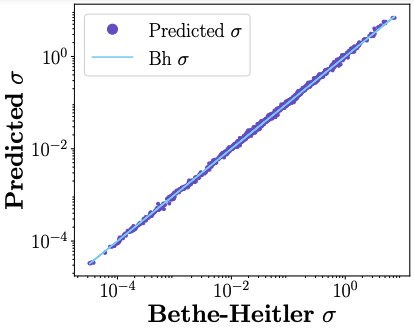}
\includegraphics[width=0.4\textwidth]{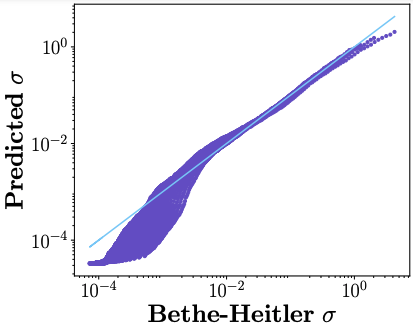}
\caption{Result of the DNN predicted BH cross section as compared to the theoretical calculation both within the test set (Top) and outside of the range data (bottom). }
\label{fig:bh}
\end{figure}

We test our trained model using pseudo-data calculated from the BH cross section. The BH cross section represents a significant portion of the full cross section at current experimental beam energies. It also has the added benefit in that it is exactly calculable in QED up to the parametrization of the elastic form factors, which for momentum transfers at current JLab DVCS experiments, are known to high precision. This means that the BH cross section is a good testing ground to benchmark our model's performance, especially outside of the range of current experimental data. We show the performance of the predicted BH as compared to the theory prediction both in the range of the data and generalizing outside of the range of the data in Figure~\ref{fig:bh}. We plot the expected value versus the predicted, meaning that if the model was performing perfectly we would see a straight line. Any non-linear behavior demonstrates predicted values that are not well understood in our model.

\subsection{Compton Form Factor Extraction}

Standard applications of supervised neural networks involve making predictions through learning the mapping from inputs to outputs in labeled datasets. We move beyond a standard supervised algorithm by attempting to extract CFFs from a single polarization channel in the cross section data. This is complicated by the fact that the data itself is not what is being predicted, rather the predicted variable is embedded in a functional composition. The loss function for this algorithm, $\mathcal{L}^{CFF}_{\theta, \phi}$ contains the functional form of the cross section parametrized by CFFs, $f_{\phi}$, with the nonlinear functions of the parameters of the network, $g_{\theta}$, embedded. 

\begin{eqnarray}
    \mathcal{L}^{CFF}_{\theta, \phi} &=& \frac{1}{2}(\sigma - f_{\phi}(g_{\theta}(x)))^2 \\
    \theta_{t+1} &=& \theta_t - \lambda \nabla_{\theta_t}\frac{1}{B}\sum_{i=0}^{B}\mathcal{L}^{CFF}_{\theta_t, \phi}(\mathbf{x}^{(i)}, \mathbf{\sigma}^{(i)}) 
    \label{eq:grad}
\end{eqnarray}

\noindent During backpropagation, when the network is updating its weights and biases through gradient descent, the gradient is much more complicated as seen in Equation \ref{eq:grad} than in a standard gradient descent problem. The loss surface is extraordinarily complex with large amounts of local minimum, this situation requires careful tuning of the the regularization parameters to attempt to smooth out the loss surface and best find the global minimum for the extracted quantity.

There exists eight leading twist CFFs that parameterize the DVCS process, and there are eight polarization configurations of lepton beam and nuclear target that are allowed by parity. This means purely from a mathematical standpoint that as long as the cross section for each polarization configuration are not degenerate in their CFF content, then one should be able to use eight equations to extract eight unknowns. This would require almost exact overlap in kinematics for all eight polarization configurations, along with significant control over systematic and statistical errors. Even then, the extraction of all eight CFFs is not well controlled. This is the importance of using artificial intelligence methods in order to tackle this extremely complicated problem, in that deep learning models provide a framework for which to perform these extractions.
%Extraction of a single CFF from the DVCS cross section. 8 CFFs from 8 observables, but we don't have data on all 8 observables and there is no theory for all 8 polarization observables. What can we learn by assuming a form of 7 CFFs and extracting one of them?

The extraction of eight CFFs from a single polarization observables is, in itself, a type of inverse problem. With a single polarization observable, or a single equation, it is not mathematically possible to extract eight unique unknowns. We make progress however, through the use of a supervised deep learning model. We can extract a single CFF from a single polarization observable using different assumptions for the other seven CFFs. In the case of a linear regression deep learning model, this is a rigorously defined problem. A full treatment of the inverse problem requires more nuanced ML methods which we outline in Section \ref{sec:vaim}.

%\begin{itemize}
%\item{Using the dvcs cross section as our loss function. CFFs are what is predicted from the NN, so the use of Machine learning here is not a standard application where the loss function is a standard distance from the predictions to the data. Instead we have to minimize the loss function which is extremely complicated.}
%
%\item{Extracting one Compton form factor. We assume a form for the rest of the 7 CFFs with 4 different assumptions: 0,1,10,and VA theory model.}
%
%\item{Unpolarized beam and unpolarized target cross section, we assume that ReH is dominant. Start with extracting ReH first.}
%
%\item{Comparison to other extractions of CFFs?}
%
%\item{Can we put constraints on what the values of the other CFFs can be? 10 changes the value of ReH a lot while 0 and 1 don't change the value of ReH very much.}
%
%\item{For each result with different assumptions for the other 7 Compton form factors we are averaging 4 runs of the NN.}
%\end{itemize}

\begin{figure}
\includegraphics[scale=0.40]{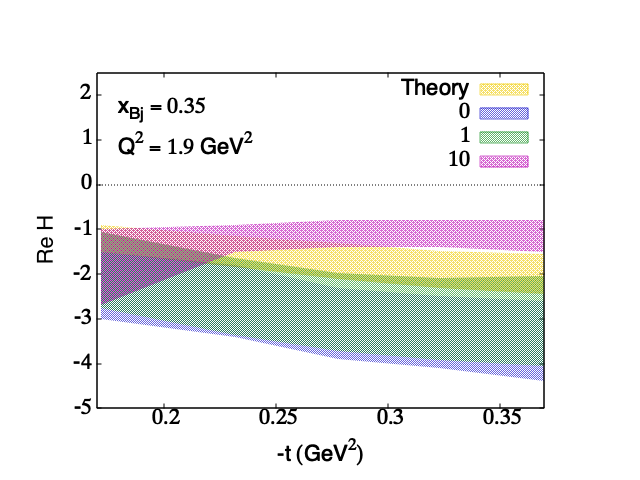}
\caption{Extraction of the CFF $\Re e \mathcal{H}(\xi,t)$ as a function of the momentum transfer $-t$ at a kinematic point $x_{Bj} = 0.35$ and $Q^{2} = 1.9$ GeV$^{2}$ , $E_e=$ 6 GeV.  Results of using various values for the other CFFs that enter the full unpolarized cross section are listed as 0, 1 and 10, as explained in the text. We also include an extraction obtained fixing the other CFFs by evaluating them using the parametrization of Refs. \cite{Goldstein:2010gu,GonzalezHernandez:2012jv,Kriesten:2021sqc}. }
\label{fig:rehvt_sgd}
\end{figure}

The unpolarized cross section is particularly sensitive to the CFF $\Re e \mathcal{H}$. We show in Figure \ref{fig:rehvt_sgd} an extraction of $\Re e \mathcal{H}$ given an assumption for the other seven CFFs which parameterize the unpolarized cross section, using values of 0,1, and 10, including a theoretical calculation whose GPD parameterization can be found in Refs. \cite{Goldstein:2010gu,GonzalezHernandez:2012jv,Kriesten:2021sqc}. It should be noticed that the value of $\Re e \mathcal{H}$ is resilient to the assumption for the value of the other seven CFFs, particularly at low-$t$, meaning that no matter the choice of the other seven CFFs, the value of $\Re e \mathcal{H}$ mostly does not change within error. This indicates that the cross section is much more sensitive to $\Re e \mathcal{H}$ at low-$t$ than at larger momentum transfer values. As the momentum transfer increases, the cross section seems to become less sensitive to $\Re e \mathcal{H}$ as the choice of the other seven CFFs starts to matter more. 

%%%%%%%%%
%%%%%%%%% Re H

%\begin{itemize}
%\item{Comparison with Rosenbluth separation method which relies on no assumption of CFFs. Using VA theory model for the 7 other CFFs in neural network extraction.}
%
%\item{Can we say something about the "model dependence" of the results from the NN? If they compare with results that have no assumption for the CFFs what does this say?}
%\end{itemize}

\begin{figure}[!ht]
\includegraphics[scale=0.40]{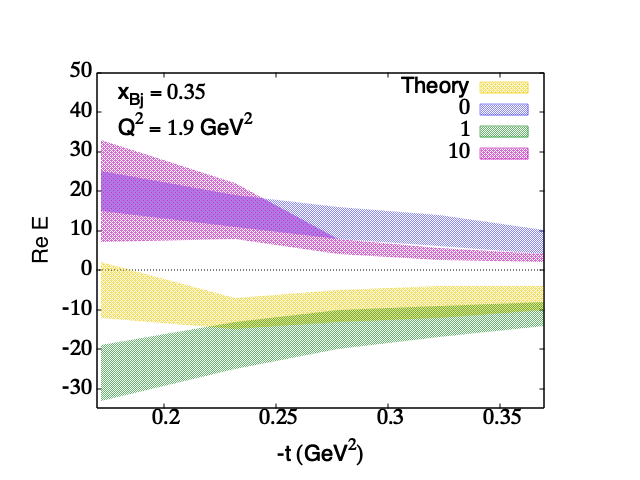}
\caption{The extraction of $\Re e \mathcal{E}$ as a function of $t$ utilizing the JLab 6 GeV data at $x_{Bj} = 0.35$ and $Q^{2} = 1.9$ GeV$^{2}$. The extracted errors are given by the random targets method.}
\label{fig:ree_solo}
\end{figure}

In Figure \ref{fig:ree_solo} we show an extraction of $\Re e \mathcal{E}$ by itself from the unpolarized cross section using  0,1,10, and model for the rest of the CFFs. One can see that the large spread in $\Re e \mathcal{E}$ demonstrates that this CFF is particularly sensitive to the value of the other CFFs. We attribute this to the fact that $\Re e \mathcal{E}$ enters the cross section through the pure DVCS cross section and the interference term, both of which their contribution is multiplied by a small coefficient, therefore there is a lot of flexibility in the fit that makes $\Re e \mathcal{E}$ particularly difficult to extract in this manner. We also note that as a function of $t$ the spread converges, this is because the contribution from $E$ grows as a function of $t$. This is in contrast with the CFF $\Re e \mathcal{H}$ which seems to have the opposite behavior. We note here that there are alternative methods such as Rosenbluth separations that we suggest for extracting the CFF $\Re e \mathcal{E}$ from the unpolarized cross section which we describe in Ref. \cite{Kriesten:2020apm}.

%%%%%%%%%
%%%%%%%%% Re E

\begin{figure}[!ht]
\includegraphics[scale=0.4]{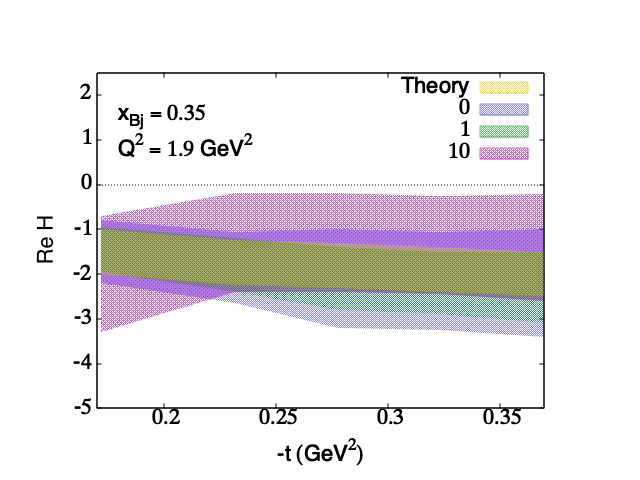}
\includegraphics[scale=0.4]{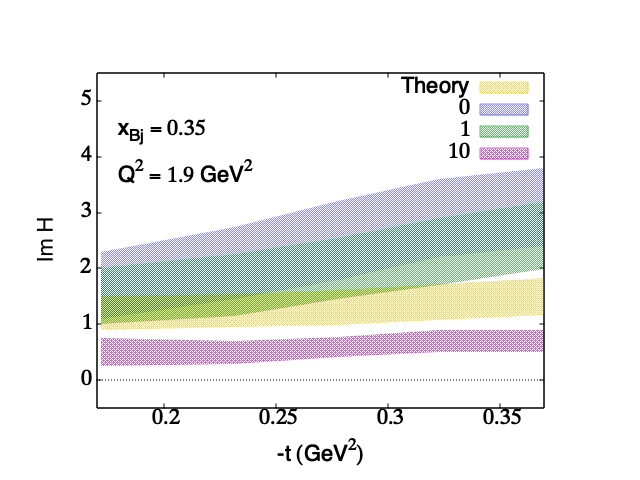}
\caption{A simultaneous extraction of $\Re e \mathcal{H}$ and $\Im m \mathcal{H}$ from the unpolarized cross section and the beam spin asymmetry at the kinematic values of $Q^{2} = 2$ GeV$^{2}$, $x_{Bj} = 0.36$, $E_{b} = 6$ GeV.}
\label{fig:rehimh} 
\end{figure}

In Figure \ref{fig:rehimh} we show the results of utilizing data from two cross sections to extract two CFFs. We use both unpolarized target cross sections (with unpolarized and polarized electron beam) to study the extraction of the two CFFs, $\Re e \mathcal{H}$ and $\Im m \mathcal{H}$ from these combined datasets. We notice that the extracted value of $\Re e \mathcal{H}$ seems to converge towards a central value regardless of the choice of the other seven CFFs, even more so than when $\Re e \mathcal{H}$ is extracted alone. The CFF $\Im m \mathcal{H}$ seems to converge more at smaller momentum transfer values and diverge at larger $t$.

\begin{figure}[!ht]
\includegraphics[scale=0.4]{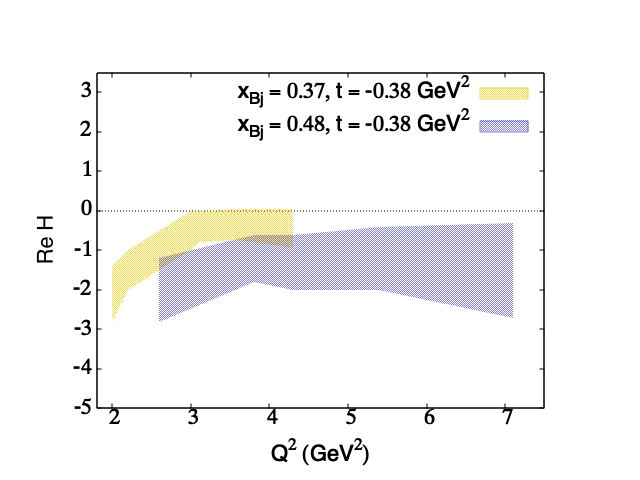}
\caption{Extracted $\Re e \mathcal{H}$ from combined JLab Hall 6 GeV and 12 GeV datasets as a function of $Q^{2}$.}
\label{fig:rehvq2_sgd} 
\end{figure}

The power of ML algorithms is through finding correlations that are prevalent through multiple datasets. Figure \ref{fig:rehvq2_sgd} demonstrates this power of a global analysis by connecting Jefferson Lab DVCS data from 6 GeV and 12 GeV to study the trends in $Q^{2}$ of the CFF $\Re e \mathcal{H}$. Studying scaling in $Q^{2}$ of the CFFs is a critical test of whether we truly understand the scattering mechanism in which these deeply virtual exclusive processes occur. We see that at low-$Q^{2}$ there is a non-linear trend in scaling of the CFF even with the larger error bars that are extracted. This is contrasted with the somewhat larger $Q^{2}$ behavior in which the dependence of the CFF on $Q^{2}$ seems to flatten. We demonstrate this at two values of $x_{Bj} = 0.37$ and $x_{Bj} = 0.48$ and at the same momentum transfer $t = -0.38$ GeV$^{2}$. It should be noted here that in order to truly understand the trends in $Q^{2}$ one would need to extend the $Q^{2}$ reach to much larger than $8$ GeV$^{2}$, a critical role is played here by the EIC and JLab at 20+ GeV, both of which provide that larger lever arm.

%%%%%%
%%%%%% 6 GeV and 12 GeV
\begin{figure}[!ht]
\includegraphics[scale=0.46]{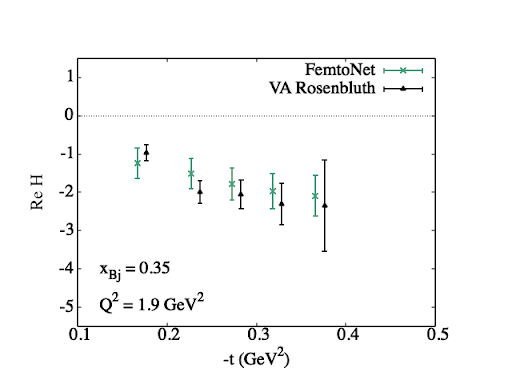}
\caption{Comparison of the extracted $\Re e \mathcal{H}$ as a function of $-t$ at a kinematic point $x_{Bj} = 0.35$ and $Q^{2} = 1.9$ GeV$^{2}$. We compare these extracted values against those obtained through a Rosenbluth separation \cite{Kriesten:2020apm}.}
\label{fig:rehvt_comp}
\end{figure}

In Figure \ref{fig:rehvt_comp} we compare the results of our ML extracted CFF $\Re e \mathcal{H}$ to our Rosenbluth separated results as first reported in \cite{Kriesten:2020apm} at the same kinematics $x_{Bj} = 0.35, Q^{2} = 1.9$ GeV$^{2}$, $-t = 0.2 - 0.4$ GeV$^{2}$, and $E_{b} = 6$ GeV. In the case of FemtoNet we show the results where the other seven CFFs are given by their model calculated results and the uncertainties are a result of the random targets uncertainty quantification technique described in section \ref{sec:error}. In the case of the Rosenbluth separation we know that there is no CFF model input as the CFFs are extracted through linear relations in the interference cross section. The fact that these two methods overlap indicates that the extraction of CFFs is most sensitive to the interference cross section. It is a subject of further study to implement these linear relations are physics constraints into the neural network architecture to improve the extraction of the CFFs. 

%%%%%

%\begin{itemize}
%
%\item{Scaling? What can we say about scaling from these measurements? We don't have a large range in $Q^{2}$.}
%\end{itemize}

%\item{ReH and ReE extracted together, what does it mean to extract 2 CFFs from 1 physical observable? 1 equation and 2 unknowns doesn't sound like it should be possible to pinpoint. But with error?}

%\begin{figure}[!ht]
%\includegraphics[scale=0.35]{figures/rehree_rehext.png}
%\includegraphics[scale=0.45]{figures/rehree_reeext.png}
%\label{fig:rehree}
%\caption{}
%\end{figure}

%\begin{itemize}
%\item{ImH extracted from $\sigma_{LU}$ data}

%\begin{figure}[!ht]
%\includegraphics[scale=0.35]{figures/imhvt_sgd.png}
%\caption{}
%\end{figure}

%\item{ReH and ImH extracted together from $\sigma_{UU}$ and $\sigma_{LU}$ data combined}

%\end{itemize}

%%%%%%%%%
%%%%%%%%% Error analysis
\section{Error Analysis}
\label{sec:error}

In our recent publication \cite{Grigsby:2020auv} we utilized dropout as an estimate of the systematic errors of the network \cite{gal2015dropout, gal2016dropout}. Dropout sets weights of a fully connected neural network layer to 0 with a probability $p$, severing connections between neurons at different layers creating a more robust learning model. The purpose is two-fold: dropout is a type of regularization technique to prevents overfitting, but it also approximates the systematic error of the ML model results of the network if left on during forward passes of the network. When the same input to a neural network produces different outputs due to different dropout masks being used in each forward pass (meaning different neuron connections are separated with each forward pass), the statistical mean of the outputs is taken as the neural network model uncertainty of the output. This uncertainty quantification technique can be shown to be mathematically equivalent to Monte Carlo sampling.

As we discussed previously, uncertainty from the data must also be taken into account so that all forms of uncertainty are propagated through our deep learning algorithm. To take into account the irreducible uncertainty from the data, we have developed a method known as ``random targets". We randomly sample from a Gaussian distribution centered at the central value of the data point allowing us to take into account the information contained in the error bars of the data. With dropout also turned on during the forward pass also while randomly sampling from within the error bars of the data, we can propagate the irreducible uncertainty from the data while also keeping track of the uncertainty introduced from the deep learning model. 

\begin{figure}[!ht]
\includegraphics[scale=0.40]{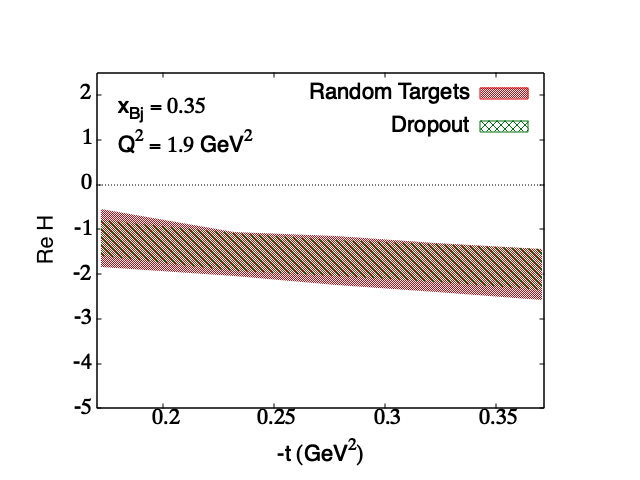}
\caption{Comparison of results when using ``random targets" method for calculation of error (red) versus predicting on the central value of the cross section with dropout turned on (green).}
\label{fig:err_comp}
\end{figure}

In Figure \ref{fig:err_comp} we show the results of using dropout to model network uncertainty alone on the central values of the cross section in blue and compare the size of the propagated error against the random targets method in red. Notice that the random targets method produces a slightly larger error bar than just the network uncertainty. This suggests that the random targets method is propagating the experimental uncertainty through the deep learning model. We also notice that the majority of the uncertainty lies in the model uncertainty. This suggests with proper tuning of the model algorithm, one should be able to extract accurate CFF predictions with reduced model uncertainty.The CFF predictions shown in the previous section all utilize the random targets technique for uncertainty propagation. There exists other forms of uncertainty quantification in the literature as was discussed in the previous sections; however, for this discussion we utilize a single propagation technique and leave a comparison of all methods to a future study.

\begin{figure}[h]
\center
{
  \includegraphics[width=0.85\columnwidth]{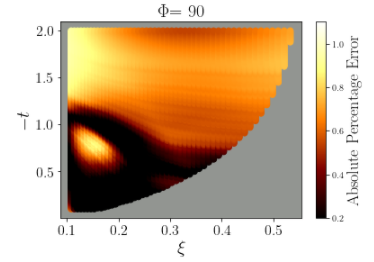}%
  \label{bh_}
}
{
  \includegraphics[width=0.85\columnwidth]{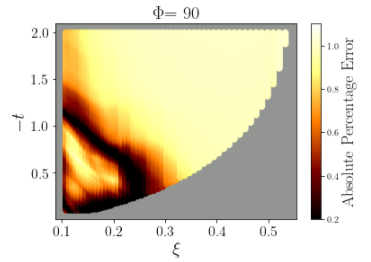}%
  \label{dvcs_}
}
\caption{(Top) The error on bh data,(Bottom) the error on DVCS data at $\Phi= 90$}
\label{fig:bh_dvcs}
\end{figure}

In Figure \ref{fig:bh_dvcs} we investigate the generalization capabilities of the deep learning model as compared to the exactly calculated Bethe-Heitler process and a calculation of the DVCS process. In the slice of kinematic phase space shown in the figure ($Q^{2} = 2$ GeV$^{2}$, $E_{b} = 6$ GeV), the experimental data lies in a small box defined between $\xi = 0.2 - 0.3$ and $-t = 0.17 - 0.4$ GeV$^{2}$ \cite{Defurne:2015kxq}. For the top figure, we take the kinematic points in $x_{Bj}, t, Q^{2}, E_{b},$ and $\phi$ and create pseudo-data for this region. We then train our neural network on this data and use it to predict in the full kinematic phase space that is shown in the figure. We compare the results of the neural network predictions against the calculated BH cross section at a given azimuthal angle $\phi = 90^{\circ}$ using the metric of absolute percentage error (APE) which is given as:

\begin{eqnarray}
    \text{APE} = \frac{|y - \sigma| }{\sigma}
\end{eqnarray}

In the bottom figure we use the exact same metric; however, we are comparing to a calculation of the DVCS cross section using a parametrization of the CFFs given in \cite{Goldstein:2010gu,GonzalezHernandez:2012jv,Kriesten:2021sqc}. It is interesting to note that the BH cross section outperforms the DVCS cross section in terms of learned generalization across the kinematic phase space. Also that there are significant trends that the deep learning model can learn when given small amount of physics information.

%%%%%%%%
%%%%%%%% Outlook
\section{Conclusions and Outlook}
\label{sec:outlooks}
CFFs have been extracted from DVES data by several groups to date, 
from earlier work reviewed in Ref.\cite{Kumericki:2016ehc}, to more recent analyses  Refs.\cite{JeffersonLabHallA:2022pnx,Shiells:2021xqo} using conventional approaches, and  Refs.\cite{Cuic:2020iwt,Moutarde:2019tqa,Grocholski:2019pqj,Kumericki:2019mgk}, using ANN-based approaches. 
In view of an upcoming broader volume of measurements from Jefferson Lab@12 GeV to the EIC, it is of the utmost importance to be able to evaluate the discrepancies among the different extraction approaches in a quantifiable way, moving forward from a lingering ``proof of concept" stage of the analysis.
We have provided a study of benchmarks for both the physical description of the cross section and the ML aspects, including a 
through treatment of uncertainty demarcating the ML generated one from the statistical experimental error.
In our ML framework we utilize a data-centric  method in which the CFF is extracted directly from unpolarized/polarized cross section data. By extracting the CFF directly from the data we minimize any bias that is introduced through fitting a parametric functional form for the CFFs. In this way, our predictions of the CFFs show the limitations of the datasets and represent exactly how much information can truly be extracted from the datasets. Information can be skewed through the fitting of a parametric form in that errors may be unfaithful to the information extracted from the datasets, especially in regions far outside of the experimental data.

Another important aspect of our analysis is that in our approach we move from the ``local" harmonics analysis used ubiquitously so far, and embrace a global perspective which doesn't need harmonics. The large number of points in $\phi$ is an artifact of a harmonics analysis, and  the harmonics analysis has yet to yield results for definitive determinations of CFFs. 

Our ML analysis supports taking into account a reduced set of data points in $\phi$ in view of the symmetry constraints of the GPD description \cite{Kriesten:2020apm}. 

%Emphasis on incorporation of physics information to accurately and efficiently predict physics observables from data. 

Benchmarking will be crucial for comparing the extracted CFFs from various groups in the literature.
Our benchmarks addressed the number and type of CFFs in unpolarized scattering, and the $Q^2$ dependence of the cross section. Future work will be extended, on one side  to include other group's extractions, and on the other to include additional features, which we list in what follows.

\subsection{Connections to Lattice QCD}
\label{sec:lqcd}
There has been recent progress made in lattice QCD to providing the Bjorken-$x$ dependence of the isovector nucleon GPDs, $H$, $E$ and $\tilde{H}$ using Large-Momentum Effective Theory (LaMET), also known as the ``quasi-PDF method''~\cite{Ji:2013dva,Ji:2014gla,Ji:2017rah,Ji:2015qla,Liu:2019urm}.
ETM Collaboration used the LaMET method to calculate both unpolarized and polarized nucleon isovector GPDs with largest boost momentum 1.67~GeV at pion mass $M_\pi \approx 260$~MeV~\cite{Alexandrou:2020zbe} with one momentum transfer.
MSULat also reported the first lattice-QCD calculation of the unpolarized and helicity nucleon GPDs with boost momentum around 2.0~GeV at the physical pion mass with multiple transfer momenta, allowing study of the three-dimensional structure and impact-parameter--space distribution~\cite{Lin:2020rxa,Lin:2021brq}.
Such lattice inputs can provide useful constraints to guide neural networks to the best determination of physical quantities using both theoretical and experimental inputs. 

Reference~\cite{Lin:2017stx} shows an example of how to use lattice tensor data to provide constraints in a global fit to the transversity PDF where experimental data is limited.
Using multiple lattice-QCD calculations, a continuum-limit extrapolation of the nucleon isovector tensor charge $g_T$ is performed to yield $1.008(56)$.
The lattice $g_T$ can be used to constrain the global-analysis fits to SIDIS $\pi^\pm$ production data from proton and deuteron targets, including their $x$, $z$ and $P_{h,\perp}$ dependence, with a total of 176 data points collected from measurements at HERMES \cite{Airapetian:2010ds} and COMPASS \cite{Alekseev:2008aa,Adolph:2014zba}. %This gives in principle 8 linear combinations of transversity TMD PDFs and Collins TMD FFs for different quark flavors, from which we attempt to extract the $u$ and $d$ transversity PDFs and the unfavored Collins FFs, together with their respective transverse-momentum widths. 
As shown in Figure.~2 of Ref.~\cite{Lin:2017stx} the lattice input has made a significant impact in constraining the transversity over a wide range of $x$, especially in the down-quark contribution. 
This is one demonstration of how the combination of experiment and precision lattice-moment inputs can greatly advance our knowledge of nucleon structure. 
One can extend the idea to combine the lattice GPD inputs with experimental data in the global fit to extract the best determination of the GPD.

\subsection{Reinforcement Learning}
\label{sec:rl}

The ultimate goal of exclusive scattering measurements is to extract physical properties of the nucleon through experimental data. Through the quantum correlation functions (GPDs) that parametrize the 3D spatial structure of the nucleon, we can connect experimental observables to quantities such as spin, mass, and pressure forces. There is a significant challenge in extracting the GPDs from the CFFs that we have access to in experiment. We propose a method of sequence to sequence reinforcement learning (RL) to point by point in $x$ recreate the GPD by systematically applying theoretical constraints, lattice QCD calculations, and phenomenological extractions of CFFs. The application of RL provides a unique synergistic opportunity to bring for the first time this technology to the study of nuclear phenomenology.

This represents the first global analysis of GPDs in the framework of machine learning. Utilizing the NN architecture to model the quantum correlation functions that we need to study QCD phenomenon and allowing us to reconstruct the $x$-dependence. Inherent to this RL study is the discussion of information theory, how much information is lost in the convolution of the GPD over the Wilson coefficient function. There have been recent attempts to answer these questions; however, RL allows us a framework in order to quantify exactly these results.

\subsection{Tackling the Inverse Problem of Extracting CFFs from Exclusive Scattering Observables}
\label{sec:vaim}

The inverse problem of systematically extracting the leading twist CFFs from polarization observables is of primary interest, which is the easiest channel to access GPD. There are eight leading twist CFFs and eight polarization configurations of beam and target allowed by parity. Neglecting the subleading quark and gluon contributions including gluon transversity CFFs to the cross section, in theory one can extract all eight leading twist CFFs from experimental measurements. It is a huge experimental undertaking to consider overlapping measurements for all polarization configurations at the exact same kinematics with controllable errors. Considering the fact that the cross section for DVCS measurements includes non-linear CFF terms, this problem becomes extraordinarily unlikely to solve in this manner. In order to gain insight to the values of the CCFs including extracted errors, we must understand how much information we can gain from a single polarization observable or the overlap of a few observables at the same kinematic point. This is an example of an inverse problem with potentially multiple solutions, since all CFFs enter into the unpolarized cross section and there is only one observable from which to determine them.

The deep neural network (DNN) methods such as Variational Autoencoder Inverse Mapper (VAIM)~\cite{Almaeen2021} is a potential venue to solve the inverse problem in CFF extraction. The VAIM is composed of an encoder and decoder NN to respectively approximate forward and backward mapping and a variational latent layer to learn the posterior parameter distribution. In previous works, the VAIM was used to determine multiple solutions to the parameterization of quark/anti-quark/gluon parton distribution functions from structure function measurements. It was found that through the use of the VAIM framework, one can reconstruct the lost information through an analysis of the latent space variables. Given cross section as input, sampling the latent layer was able to reconstruct the lost information and as a consequence determine the corresponding parameter distribution of the parton distribution function. 

We propose a similar goal in the context of DVCS. Treating the experimental observable as an equation that is parametrized by CFFs, we can recast the VAIM framework to study the effect of extracting CFFs and multiple solutions. Through the data augmentation method that we describe in this work, we will be able to then propagate experimental error into the VAIM analysis to quantify the uncertainty associated with CFF parameterization, which may prove more beneficial to analyzing exclusive scattering experiments. 
% through an ensembling method similar to the replica method (cite...). This type of architecture will also allow us to explore other methods of uncertainty quantification which may prove more beneficial to analyzing exclusive scattering experiments. 

\subsection{Applications of UQ}
\label{sec:uq}

With increased use of neural networks in phenomenology, an essential question is 'how confident we should be in the DNN model's predictions?' Confidence in neural network predictions comes from a thorough understanding of the model, its ability to make decisions, and a controllable uncertainty estimation on those decisions. Standard applications of statistical inference in data analysis have been used to extract information from experimental data; however, with the incoming wave of high impact experimental measurements from a vast array of exclusive scattering channels it becomes critical to establish a computational framework that can handle this large system of data as well as understand complex correlations between the datasets. Uncertainty quantification is a field of study which seeks to understand the answers to these questions. It is key to understand the separation of error into categories of systematic to the network method, reducible model uncertainty (epistemic), and propagated from the experimental data, non-reducible data uncertainty (aleatoric). By separating out the epistemic error, we can work to reduce the network error through UQ techniques and generate a controllable, propagated error from the data into our extracted observable.

\vspace{0.5cm}

\acknowledgements
We thank our extended FemtoNet collaboration, and Charles Hyde and Michelle Kuchera for discussions.
This work was funded by DOE grants DE-SC0016286 (S.L.), PHY 1653405 and  Research  Corporation  for  Science  Advancement through the Cottrell Scholar Award (H.L.), the  
SURA Center for Nuclear Femtography (B.K., S.L., J.G., J.H., M.A., Y.L.), and a PhD scholarship from Jouf University, Saudi Arabia (M.A.).

\bibliography{BIB}
\end{document}